\documentclass[journal]{IEEEtran}
\usepackage{epsfig}
\usepackage{cite,url}
\usepackage{graphicx,color}
\usepackage[cmex10]{amsmath}
\usepackage{amsfonts,amssymb,mathrsfs,wasysym,stackrel}
\usepackage{algorithmic}
\usepackage{algorithm}
\usepackage{latexsym}
\newcommand{\ra}{\rightarrow}

\newcommand{\bo}{\textbf}
\newcommand{\ita}{\textit}
\newcommand{\F}{\mathcal{F}}

\newcommand{\SA}{\mathcal{S}}
\newcommand{\G}{\mathcal{G}}

\newcommand{\E}{\mathcal{E}}
\newcommand{\N}{\mathcal{N}}
\newcommand{\V}{\mathcal{V}}
\newcommand{\R}{\mathcal{R}}

\newtheorem{thm}{Theorem}

\newtheorem{remark}[thm]{Remark}

\begin{document}

\title{Asynchronous Gossip for Averaging and Spectral Ranking}

\author{Vivek~S.~Borkar, Rahul~Makhijani and~Rajesh~Sundaresan
\thanks{V.~S.~Borkar is with the Department of Electrical Engineering, IIT Bombay, Powai, Mumbai 400076, India.}
\thanks{R.~Makhijani was with IIT Bombay when this work was done.}
\thanks{R.~Sundaresan is with the ECE Department, Indian Institute of Science, Bangalore 560012, India.}
\thanks{Portions of this work were presented at the 50th Allerton Conf.\ on Communication, Control, and Computing, Oct.\ 1-5, 2012, Monticello, IL, \cite{bormak} and at the Workshop on Information Theory and Applications, San Diego, Feb.\ 10-15, 2013 \cite{bormaksun}. Research of V.~S.~Borkar was supported in part by a J.~C.~Bose Fellowship and a grant ``Distributed Computation for Optimization over Large Networks and High Dimensional Data Analysis'' from the Department of Science \& Technology, Government of India. R.~Sundaresan was supported by the Indo-US Science and Technology Forum Fellowship and by the US National Science Foundation under grant CCF-1017430. This author thanks the Coordinated Sciences Laboratory, University of Illinois at Urbana-Champaign, for its hospitality during the course of this work.}
}

\maketitle

\begin{abstract}
We consider two variants of the classical gossip algorithm. The first variant is a version of asynchronous stochastic approximation. We highlight a fundamental difficulty associated with the classical asynchronous gossip scheme, viz., that it may not converge to a desired average, and suggest an alternative scheme based on reinforcement learning that has guaranteed convergence to the desired average. We then discuss a potential application to a wireless network setting with simultaneous link activation constraints. The second variant is a gossip algorithm for distributed computation of the Perron-Frobenius eigenvector of a nonnegative matrix. While the first variant draws upon a reinforcement learning algorithm for an average cost controlled Markov decision problem, the second variant draws upon a reinforcement learning algorithm for risk-sensitive control. We then discuss potential applications of the second variant to ranking schemes, reputation networks, and principal component analysis.
\end{abstract}

\begin{IEEEkeywords}
gossip algorithm; asynchronous stochastic approximation; Poisson equation; learning; Perron-Frobenius eigenvector; ranking
\end{IEEEkeywords}

\section{Introduction}
\label{sec:Introduction}

The so called `gossip algorithm' has been a popular framework for formulation of distributed schemes for in-network computation. Some of the problems that have been addressed in this framework are consensus or averaging, finding the maximum or minimum, computation  of separable functions, etc., an excellent survey of which can be found in \cite{Shah}. A more recent application is a scheme for clock synchronization \cite{Schenato}. In this article, we address two distinct albeit related issues regarding gossip schemes. In the first, we consider the plain vanilla gossip for computing a desired weighted average of the initial values and observe that a natural asynchronous sampling based version may not exhibit the desired asymptotic behavior. Having analyzed this situation, we propose an alternative scheme based on analogous constructs from reinforcement learning for approximate dynamic programming. At the expense of some added complexity, this scheme has guaranteed convergence to the desired value. Our second aim here is to broach yet another class of problems that seems amenable to gossip based computation, viz., computing the Perron-Frobenius eigenvector, i.e., the positive eigenvector of an irreducible nonnegative matrix. This has been a popular means of ranking and related evaluative exercises, particularly with the success of PageRank \cite{Langville}. The scheme proposed here is simpler and more amenable to distributed implementation than the more general schemes for principal component analysis analyzed in \cite{Oja}, \cite{Kempe}.

The connecting thread between the two problems, asynchronous gossip for averaging and spectral ranking via computation of the Perron-Frobenius eigenvector, is that both draw upon reinforcement learning algorithms for controlled Markov chains. The former draws upon an algorithm for the average cost problem while the latter draws upon an algorithm for  risk-sensitive control. The dynamic programming equation and learning algorithm for the latter can be viewed as multiplicative counterparts of the additive structure in the former.

Section \ref{sec:plain-vanilla-gossip} describes the basic gossip scheme for averaging and its stochastic approximation variant for distributed in-network computation, and points out the pitfalls thereof. Section \ref{sec:rl} motivates and analyzes the reinforcement learning algorithm which works around the pitfalls to guarantee desired convergence properties. Section \ref{sec:gossip-csma} considers a naturally asynchronous situation arising in a wireless setting and uses a protocol of Jiang and Walrand \cite{Jiang-Walrand}, \cite{Jiang} for distributed routing. Section \ref{subsec:multihop} describes an important variant, viz., a multihop version. Section \ref{subsec:importance-sampling} sketches a plausible conditional importance sampling scheme for accelerating convergence. With section \ref{sec:pca}, we analyze the second problem described above -- that of computing the Perron-Frobenius eigenvector of an irreducible nonnegative matrix in a distributed fashion. Section \ref{sec:applications} describes specific applications to ranking schemes, reputation networks, and principal component analysis.

\section{Plain vanilla gossip}
\label{sec:plain-vanilla-gossip}

Suppose that there are $d$ agents in a network. Agent $i$, $1 \leq i \leq d$, is endowed with an initial value $x_i(0) \in \R$. Write $x(0) = (x_1(0), x_2(0), \ldots, x_d(0))^T$. The agents successively exchange information and compute to arrive at a desired consensus value, which is taken to be a certain convex combination of the initial values held by the agents. Specifically, we take $P = [[p(i,j)]] \in \R^{d\times d}$ to be a given irreducible and aperiodic stochastic matrix with $\eta = [\eta_1, \ldots, \eta_d]^T$ its unique stationary distribution. The desired consensus value is then $\eta^T x(0) = \sum_{i=1}^d \eta_i x_i(0)$. Each agent's goal is to settle at this consensus value. The stochastic matrix $P$ typically arises from some local neighborhood structure and $\eta_i$ denotes the desired weight attached to the value of agent $i$.

The basic gossip algorithm \cite{DeGroot}, when agents have access to their neighbors' latest values, is a successive averaging scheme of the form
\begin{equation}
  x(n+1) = Px(n), \ n \geq 0, \label{gossip}
\end{equation}
which computes successive averages $x(n+1) \in \R^d$ of the previous iterate $x(n) \in \R^d$ with respect to the stochastic matrix $P$, beginning with the initial data vector $x(0)$. As $P$ is irreducible and aperiodic, this leads to the convergence \cite[Th.~4.9]{Levin-Peres-Wilmer}
\begin{displaymath}
  x(n) \ra \left( \eta^T x(0) \right) \textbf{1}, \mbox{ as } n \rightarrow \infty
\end{displaymath}
where $\eta$ is the unique stationary distribution for $P$ and $\textbf{1} := [1, 1, \ldots, 1]^T$ is the vector of $d$ ones. Convergence is to the desired consensus value. An `incremental' version is
\begin{displaymath}
  x(n+1) = (1 - a)x(n) + a P x(n), \ n \geq 0,
\end{displaymath}
where $a \in (0, 1]$ is a parameter that modulates the emphasis put on others' opinions as opposed to one's own evaluation.

In network applications, one often needs to consider a \textit{stochastic approximation} \cite{BorkarBook} version of the above wherein at each time $n$, the agent $i$ polls a neighbor $j$ according to probability $p(i,j)$ and `pulls' the latter's data $x_j(n)$ for averaging. The new recursion is
\begin{equation}
  \label{sagossip}
  x_i(n+1) = (1 - a)x_i(n) + ax_{\xi_i(n+1)}(n), \ 1 \leq i \leq d, \ n \geq 0,
\end{equation}
where $\xi_i(n+1)$ is generated with probability $p(i, \cdot )$ independently of all other random variables realized till time $n$. By adding and subtracting the one step conditional expectation of the last term, (\ref{sagossip}) can be written as
\begin{displaymath}
  x(n+1) = (1 - a)x(n) + a\Big(Px(n) + M(n+1)\Big),  \ n \geq 0,
\end{displaymath}
where
\begin{equation}
  \label{eqn:martingale-noise}
  M(n+1) = [M_1(n+1), \ldots, M_d(n+1)]^T
\end{equation}
is defined by
\begin{displaymath}
  M_i(n+1) := x_{\xi_i(n+1)}(n) - \sum_jp(i, j)x_j(n).
\end{displaymath}
The sequence $\{ M(n) \}_{n \geq 1}$ is a martingale difference sequence with respect to the family of $\sigma$-fields
\[
  \mathcal{F}_n := \sigma(\xi_i(m), x(m), m \leq n, 1 \leq i \leq d), \quad n \geq 0.
\]
The recursion (\ref{sagossip}) then becomes an instance of the `constant stepsize' version of the classical Robbins-Monro scheme for stochastic approximation:
\begin{equation}
  x(n+1) = x(n) + a\Big(h(x(n)) + M(n+1)\Big), \ n \geq 0, \label{SA}
\end{equation}
for a Lipschitz $h : \R^d \mapsto \R^d$. Under reasonable conditions (see \cite[Ch.~9]{BorkarBook}), the iterate in (\ref{SA}) tracks the asymptotic behavior of its limiting ordinary differential equation (in a sense that is made precise in \cite[Ch.~9]{BorkarBook}, see this paper's Appendix)
\begin{equation}
  \dot{x}(t) = h(x(t)), \ t \geq 0, \label{fluid}
\end{equation}
which for us is the linear system
\begin{equation}
  \dot{x}(t) = (P - I)x(t), \ t \geq 0. \label{ode}
\end{equation}
Here $I \in \R^{d\times d}$ is the identity matrix. Since $P$ is stochastic,  $\textbf{1} = [1, 1, \ldots,1]^T$ is the right Perron-Frobenius eigenvector of $P$. It is easy to see that $x(t)$, the solution to (\ref{ode}), converges to $(\eta^Tx(0))\textbf{1}$, which depends on the initial condition $x(0)$ \cite[Th.~20.1]{Levin-Peres-Wilmer}. The convergence rate of this (hence of (\ref{SA})) as well as the convergence rate of the original discrete scheme (\ref{gossip}) are dictated by the eigenvalue of $P$ with the second highest absolute value. We shall refer to this as the `second eigenvalue' henceforth. This has prompted a lot of analysis and algorithms for minimizing the second eigenvalue, \textit{ipso facto} maximizing the rate of convergence \cite{Boyd1}, \cite{Shah}.

The stochastic approximation version already introduces `noise', as we are replacing an averaging operation by a sample picked according to the averaging probability weights. An additional complication arises when the implementation is \textit{asynchronous} wherein,
\begin{itemize}
  \item not all components of $x(n)$ are updated concurrently,
  \item the agents may update at differing frequencies.
\end{itemize}
This is often the case in wireless systems where link activation constraints disallow simultaneous transmission of certain links. This leads to several nontrivial complications not present in the deterministic versions (\ref{gossip}) or (\ref{ode}). Even though (\ref{gossip}) and (\ref{ode}) converge to the unique (desired) limit $(\eta^T x(0)) \textbf{1}$ for any initial condition $x(0)$, the same may not be true of the stochastic case (\ref{SA}). We do, however, have the following. Let $Y_n(i) := I\{i\mbox{th component is updated at time } n\}$ where $I\{A\}$ is the indicator of an event $A$. The asynchronous updates are given by the following modification of (\ref{sagossip}):
\begin{eqnarray}
  \lefteqn{ x_i(n+1) = \left( 1 - (a Y_n(i)) \right) x_i(n) + (a Y_n(i)) x_{\xi_i(n+1)}(n),} \nonumber \\
  && \hspace*{1.3in}   \mbox{ for } 1 \leq i \leq d \mbox{ and } n \geq 0.   \label{sagossipasync}
\end{eqnarray}

\begin{thm}
\label{thm:convergence-constant-gain}
Consider the iterates in (\ref{sagossipasync}). Suppose that for each $i$, $\liminf_{n \rightarrow \infty} n^{-1}\sum_{k=0}^{n-1} Y_n(i) > 0$ almost surely. Then $x(n)$ converges almost surely to a constant multiple of $\textbf{1}$.
\end{thm}

\begin{IEEEproof}
Since each component of every iterate is a convex combination of components of the previous iterate, boundedness of the iterates is obvious. We next argue that the iterates converge almost surely (a.s.) to the set $A := \{c\bo{1} : c \in \R\}$.

Let $x^* := c^*\bo{1}$ for some $c^* \in \R$. Then
\begin{eqnarray*}
  x_i(n+1) - x_i^* &=& \left( 1 - (aY_n(i)) \right) (x_i(n) - x_i^*) \\
  & & + ~ ( aY_n(i) ) \left( x_{\xi_i(n+1)}(n) - x_{\xi_i(n+1)}^* \right).
\end{eqnarray*}
Defining the `span seminorm' $\|x\|_{sp} := \max_ix_i - \min_ix_i$, we observe that
\begin{displaymath}
  \|x(n + 1) - x^*\|_{sp} \leq \|x(n) - x^*\|_{sp}.
\end{displaymath}
Thus $\|x(n) - x^*\|_{sp}$ converges.

Suppose it converges to a strictly positive number. For any $i$, the event $\{ Y_n(i) = 1 \mbox{ infinitely often (i.o.)} \}$ occurs a.s. For any $i, j$ such that $p(i, j) > 0$, $\{\xi_i(n) = j\}$ i.o.\ a.s.\ on the set $\{Y_n(i) = 1$ i.o.$\}$, by the conditional Borel-Cantelli lemma \cite[p.\ 96,\ Cor.\ 5.29]{Breiman}.
Since the latter has probability $1$ for all $i$, it follows that both
\begin{align*}
  & \cup_i\{ Y_n(i) \hspace*{-0.05in} = \hspace*{-0.05in} 1, i \hspace*{-0.03in} \in \hspace*{-0.03in} \arg \min_k x_k(n), \xi_i(n+1) \hspace*{-0.03in} \notin \hspace*{-0.03in} \arg \min_k x_k(n)\} \\
  & \cup_i \{Y_n(i) \hspace*{-0.05in} = \hspace*{-0.05in} 1, i \hspace*{-0.03in} \in \hspace*{-0.03in} \arg \max_k x_k(n), \xi_i(n+1) \hspace*{-0.03in} \notin \hspace*{-0.03in} \arg \max_k x_k(n)\}
\end{align*}
occur i.o., a.s., on the set $\{\lim_{n\uparrow\infty}\|x(n) - x^*\|_{sp} > 0\}$, and it further follows that $\|x(n) - x^*\|_{sp}$ must strictly decrease i.o.\ as long as it is nonzero. Now note that the laws of
\[
  [(x(N + m), Y_{N+m}, \xi(N+m+1)), m \geq 0], N \geq 0,
\]
are tight as probability measures on $(\mathcal{R}^d \times \{0,1\}^d \times \V^d)^{\infty}$, where $\V = \{1,\ldots,d\}$ is the set of agents. Let $(\check{x}(n), \check{Y}_n, \check{\xi}(n+1)), n \geq 0$, denote a limit in law thereof as $N\uparrow\infty$. By the foregoing, $\|\check{x}(n) - x^*\|_{sp}$ must equal a possibly random constant $C \geq 0$ independent of $n$. We now make the crucial observation that $\{\check{x}(n)\}$ satisfies the same stochastic dynamics as $\{x(n)\}$. This is because $\liminf_{n \rightarrow \infty} n^{-1} \sum_{k=0}^{n-1} \check{Y}_k(i)$ continues to be strictly positive for each $i$, and the samplings $\check{\xi}_i(n+1)$ continue to be distributed according to $p(i,\cdot)$ independent of other samplings. By the foregoing, $\|\check{x}(n) - x^*\|_{sp}$ must decrease i.o.\ on $\{C > 0\}$, a contradiction unless $C = 0$. Hence $\|x(n) - x^*\|_{sp}$ must converge to zero a.s. This ensures a.s. convergence to $A$.

We also have $\|x(n + 1) - x^*\|_{\infty} \leq \|x(n) - x^*\|_{\infty},$ i.e., the max-norm distance of $x(n)$ from \textit{each} $x^* \in A$ is monotone. This ensures that the convergence is to a single, possibly random, point in $A$; otherwise $\|x(n) - x^*\|_{\infty}$ cannot simultaneously decrease for two distinct choices of $x^* \in A$.
\end{IEEEproof}

\begin{remark}
\label{rem:StocApprox}
If we consider a decreasing stepsize schedule $\{a(n)\}$ with $\sum_n a(n) = \infty, \sum_na(n)^2 < \infty$, then the proof is in fact simpler, because the iterates have the same a.s.\ asymptotic limit set as the o.d.e., viz., $A$ (see Appendix). The final argument in the last paragraph of the above proof then ensures convergence to a constant multiple of $\textbf{1}$. A similar result is established  in \cite{Huang}.
\end{remark}

Constant stepsize schemes have higher fluctuations in general; so we introduce a parallel averaging scheme at each node as follows, in order to reduce variance:
\begin{equation}
z(n+1) = z(n) + \frac{1}{n+1}\left(x(n+1) - z(n)\right), \ n \geq 0. \label{average}
\end{equation}
This leads to a more graceful convergence.

\begin{figure}[tb]
\centering
\includegraphics[width=3.49in]{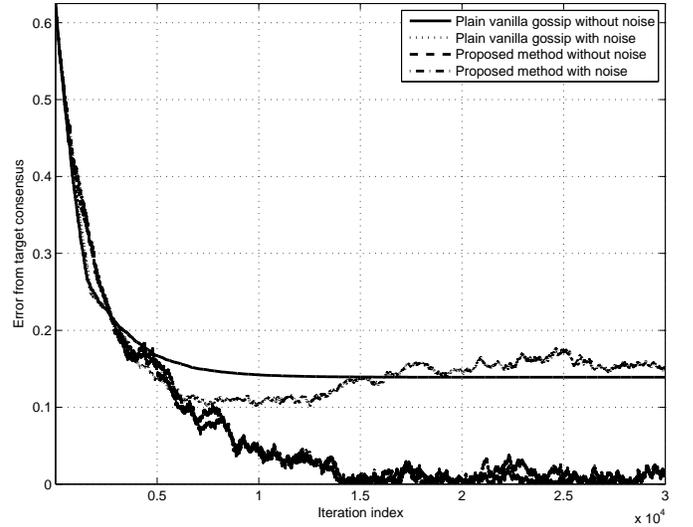}
\caption{Plain vanilla gossip reaches wrong consensus value.}
\label{fig:wrongconsensus}
\end{figure}

Pleasing as the result of Theorem \ref{thm:convergence-constant-gain} may be, it falls short of our target. It does ensure \textit{consensus}, i.e., convergence to a common value, but not to the \textit{desired} common value $\eta^T x(0)$  which is the stationary average.  We also consider the case of `noisy measurements' wherein $x_{\xi_i(n+1)}(n)$ above gets replaced by $x_{\xi_i(n+1)}(n) + W_i(n+1)$ for any i.i.d.\ zero mean noise $\{W_i(n)\}$ with finite and positive variance. In this case, the constant stepsize scheme does not even converge. These behaviors can be seen in Figure \ref{fig:wrongconsensus}.

{\em Simulation -- Description and discussion}: Convergence to a wrong consensus value can be seen even on the simplest of networks with just two nodes, but node 2 updates twice as fast as node 1. This could represent the case when node 2 receiver sees more ambient interference than node 1 receiver. The stochastic matrix is such that node 1 samples node 2 with probability 0.3 and node 2 samples node 1 with probability 0.5. The initial value $x(0)$ is $[0 \quad 1]^T$. The smooth curve in Figure \ref{fig:wrongconsensus} represents plain vanilla gossip. The plotted errors are the supremum norms of errors $x(n) - (\eta^T x(0)) {\bf 1}$. Consensus is reached (not shown in figure) at 0.2306, which is a value different from the target stationary average $\eta^T x(0) = 0.3750$. The dotted line hugging the smooth curve is for plain vanilla gossip when the data is corrupted by additive white Gaussian noise of variance 0.25. This curve hovers around the reached (but incorrect) consensus, but does not converge. The two curves at the bottom are the subject of the next section.

\section{A reinforcement learning twist}
\label{sec:rl}

The foregoing discussion prompts us to consider a different scheme to ensure convergence to the desired stationary average. The scheme is motivated by reinforcement learning algorithms for an average cost Markov decision problem \cite{ABB}. This is based on the discrete Poisson equation
\begin{equation}
  V = PV + x(0) - \beta \textbf{1}, \label{DP}
\end{equation}
which is to be solved for the pair $V(\cdot) \in \R^d, \beta \in \R$. Under an irreducibility hypothesis on $P$, (\ref{DP}) has a solution $(V^*(\cdot), \beta^*)$. Moreover, $V^*$ is specified uniquely modulo an additive scalar constant, whereas $\beta^*$ is characterized uniquely as the optimal cost: $\beta^* = \eta^Tx(0)$, where once again $\eta$ is the stationary distribution for $P$. The quantities $V^*$ and $\beta^*$ can be computed by the \textit{relative value iteration} scheme, given by
\begin{displaymath}
  V(n+1) = P V(n) + x(0) - V_{i_0}(n) \textbf{1}, \ n \geq 0,
\end{displaymath}
where $i_0$ is a fixed state. It can be shown that $V(n) \rightarrow V^*$ and $V_{i_0}(n) \rightarrow \beta^*$, where $V^*$ is the unique solution of (\ref{DP}) satisfying $V^*_{i_0} = \beta^*$. See \cite{Puterman} for these and related facts for relative value iteration on the more general controlled Markov chain, i.e.,
\[
  V_i(n+1) = \min_u \left[ \sum_j P(j | i, u) V_j(n) + c(i,u) - V_{i_0}(n)\right],
\]
where $u$ is a control parameter, $P(j|i,u)$ is the transition probability matrix for the controlled Markov chain, and $c(i,u)$ is the more general cost function. This more general case simplifies to our special case when the control parameter is degenerate and $c(i,u) \equiv x(0)$.

A stochastic approximation version of our special case of relative value iteration can be given along the lines of Abounadi et al. \cite{ABB} as
\begin{equation}
  y_i(n+1) = (1 - a)y_i(n) + a\left(y_{\xi_i(n+1)}(n) + x_i(0) - y_{i_0}(n)\right). \label{RL}
\end{equation}
It is proved in \cite[Th.~3.5]{ABB} that $y(n) \rightarrow V^*, y_{i_0}(n) \rightarrow \beta^*$, a.s.\ when the stepsize schedule satisfies the conditions of Remark \ref{rem:StocApprox}. For sake of completeness, we provide a short proof here for the case of constant stepsize. (See also Theorem \ref{thm:finite-time-analysis} later in this section).

\begin{thm}
\label{thm:RVI}
The iterates (\ref{RL}) satisfy
\begin{eqnarray*}
\limsup_{n\uparrow\infty}E[\|y(n) - V^*\|^2] &=& O(a) \\
\limsup_{n\uparrow\infty}E[\|y_{i_0}(n) - \beta^*\|^2] &=& O(a).
\end{eqnarray*}
\end{thm}

\begin{IEEEproof}
The limiting o.d.e.\ is a linear system of the form $\dot{x}(t) = Ax(t) + b$, where $A = P - I - \bo{1}e_{i_0}^T$, $e_i$ being the unit vector in $i$th coordinate direction, and $b = x(0)$. By a result of Brauer \cite[Th.~2.8]{Brauer}, it follows that the eigenvalues of $A$ are precisely $-1$ and $\lambda - 1$ where $\lambda$ ranges over the eigenvalues of $P$ other than $1$. Since the latter are strictly less than $1$ in absolute value, this is a stable linear system. The rest is routine from the `o.d.e.' analysis of stochastic approximation summarized in the Appendix's Theorems \ref{appthm:convergence-constant-a} and \ref{appthm:bounded-constant-a}.
\end{IEEEproof}

\begin{remark}
Note that the second claim shows that the aim of computing $\beta^* = \eta^T x(0)$ (if approximately) is achieved. See Remark \ref{rem:modifiedRVI} and the subsequent discussion.
\end{remark}

{\em Further discussion of simulation results}: Our numerical experiments showed significant reduction in error when (\ref{RL}) was used in place of (\ref{sagossip}). The additional measurement noise $\{W(n)\}$ introduced above does not affect the conclusions. The two curves at the bottom in Figure \ref{fig:wrongconsensus} depict the error in iterates (\ref{RL}) with and without noise. The setting is the same as the setting for the top two curves of Figure \ref{fig:wrongconsensus} and was described in the previous section.

One can go a step further than in Theorem \ref{thm:RVI} and do finite time analysis.

\begin{thm}
\label{thm:finite-time-analysis}
Let $e(n) := y(n) - V^*$ denote the error in the iterates in (\ref{RL}) with $0 < a < 1$, where $V^*$ solves (\ref{DP}) with $V^*_{i_0} = \beta^*$. Let $A = P - I - \bo{1}e_{i_0}^T$. Then the following hold.
\begin{itemize}
  \item[(i)] The spectral radius $\rho(I + aA) < 1$.
  \item[(ii)] $E[e(n)] = (I + aA)^n e(0)$, and this expected error converges to zero exponentially fast.
  \item[(iii)] With $||I+aA||$ denoting the operator norm of the matrix $I+aA$, the sum
  $$\sum_{k=0}^{\infty} || (I+aA)^k || =: C < \infty.$$
  \item[(iv)] The error concentrates in the following sense. For all $K > 0$ and all $n \geq 1$, we have
  \[
    \hspace*{-.2in} \Pr \left\{ || e(n) - E[e(n)] || \geq K a \right\} \leq 2d \cdot e^{- K^2 / \left( 4 C d^2 ||x(0) ||_{\infty} \right)}.
  \]
\end{itemize}
\end{thm}

\begin{IEEEproof}
(i) The eigenvalues of $A = P - I - \textbf{1} e_{i_0}^T$ are $-1$ and $\lambda - 1$ where $\lambda$ ranges over the eigenvalues of $P$ other than $1$; see justification in the proof of Theorem \ref{thm:RVI}. The eigenvalues of $I + aA$ are therefore $1-a$ and $1 - a(1 - \lambda)$ where $\lambda$ ranges over the eigenvalues of $P$ other than $1$. Since $0 < a < 1$, we have $|1 - a| < 1$. Furthermore, using $|\lambda|<1$, we have
\begin{eqnarray*}
  |1 - a(1 - \lambda)|^2 & \leq & (1 - a)^2 + a^2 |\lambda|^2 + 2a(1-a)|\lambda| \\
  & < & (1 - a)^2 + a^2 + 2a(1-a) ~ = ~ 1.
\end{eqnarray*}
Thus the spectral radius $\rho(I + aA) < 1$.

(ii) Observe that for $n \geq 0$, we have
\[
  y(n+1) = (I + aA) y(n) + ax(0) + aM(n+1),
\]
where $M(n+1)$, defined in (\ref{eqn:martingale-noise}), is a bounded martingale difference sequence. (We can take the bound on each component to be $2||x(0)||_{\infty}$). Since $V^*$ satisfies (\ref{DP}), we also have
\[
  V^* = (I + aA) V^* + ax(0).
\]
From these two equations, we get
\[
  e(n+1) = y(n+1) - V^* = (I+aA) e(n) + a M(n+1).
\]
Iterating this, for all $n \geq 0$, we have
\begin{equation}
  e(n) = (I+aA)^n e(0) + a \sum_{k=1}^n (I + aA)^{n - k}M(k). \label{eqn:error}
\end{equation}
The martingale-difference term disappears when an expectation is taken. Since $\rho(I + aA) < 1$, the second statement follows .

(iii) This statement easily follows from Gelfand's formula $\rho(I + aA) = \lim_{k \rightarrow \infty} || (I + aA)^k ||^{1/k}$ and $\rho(I + aA) < 1$.

(iv) From (\ref{eqn:error}), the event $||e(n) - E[e(n)] || \geq Ka$ is the same as $||\sum_{k=1}^n (I+aA)^{n-k} M(k) || \geq K$. We now bound the probability of this event. For $1 \leq k \leq n$, we have
\begin{eqnarray*}
  ||(I+aA)^{n-k} M(k)|| & \leq & ||(I+aA)^{n-k}|| \cdot ||M(k)|| \\
   & \leq & ||(I+aA)^{n-k}|| \cdot \left( 2 d ||x(0)||_{\infty} \right).
\end{eqnarray*}
It follows that for any $i$, the $i$th component martingale difference sequence is bounded by
\[
  |((I+aA)^{n-k} M(k))_i| \leq ||(I+aA)^{n-k}|| \cdot \left( 2 d ||x(0)||_{\infty} \right).
\]
Apply McDiarmid's inequality \cite[Lem. 4.1]{McDiarmid} to get
\begin{eqnarray*}
  \lefteqn{ \Pr \left\{ \left| \sum_{k=1}^n \left( (I + aA)^{n-1} M(k) \right)_i \right| \geq \frac{K}{\sqrt{d}} \right\} } \\
    & \leq & 2 e^{ - (K/\sqrt{d})^2 / (4 d ||x(0)||_{\infty} \sum_{k=1}^n ||(I+aA)^{n-k}|| ) } \\
    & \leq & 2 e^{ - K^2 / (4 d^2 ||x(0) ||_{\infty} \sum_{k=0}^\infty ||(I+aA)^k|| ) } \\
    & = & 2 e^{ - K^2 / (4 C d^2 ||x(0) ||_{\infty}) }.
\end{eqnarray*}
Finally, $||\sum_{k=1}^n (I+aA)^{n-k} M(k) || \geq K$ implies that there is a component $i$ such that $\left| \sum_{k=1}^n \left( (I+aA)^{n-k} M(k) \right)_i \right| \geq K/\sqrt{d}$. The result now follows from the union bound.
\end{IEEEproof}

\begin{remark}
\label{rem:modifiedRVI}
As pointed out in \cite{ABB}, we can replace $y_{i_0}(n)$ above by $f(y(n))$ for any $f: \R \mapsto \R$ satisfying $f(\textbf{1}) = 1$ and $f(\textbf{x} + c\textbf{1}) = f(\textbf{x}) + c$ for $c \in \R$.
\end{remark}

\begin{figure}[bt]
\centering
\includegraphics[width=3.49in]{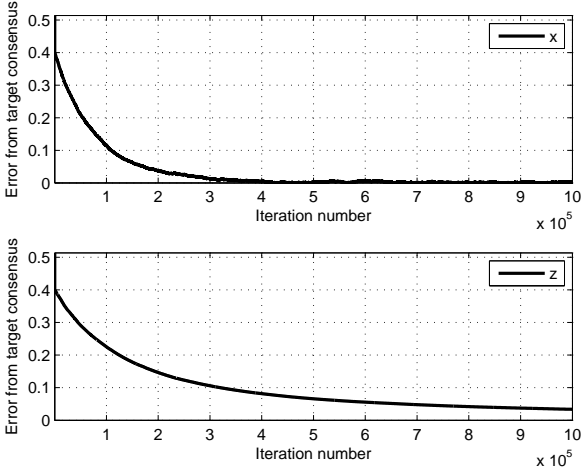}
\caption{Relative value iteration on an Erd\"os-R\'enyi graph with 100 nodes.}
\label{fig:ER-RVI-noise}
\end{figure}

One important issue with (\ref{RL}) is that the $i_0$th component of the iteration has to be broadcast to all nodes. This can be done by another gossip algorithm as in \cite[Ch.~3]{Shah} on a faster time scale. Alternatively, we can replace it by a suitable weighted average of the $x_i(n)$'s that is computed in a distributed manner, again by a gossip algorithm akin to the one above, on a faster time scale as follows: Let $R = [[r(i,j)]]_{i,j \in \V}$ be an irreducible stochastic matrix compatible with the graph $\G$ (whose nodes are $\{ 1, \ldots, d\}$ and whose edges are all pairs for which $p(i,j) > 0$), with $\kappa := [\kappa_1, \ldots, \kappa_d]^T$ as its unique invariant probability vector. Then $f(y(n)) := \kappa^T y(n)$ is a possible replacement for $y_{i_0}(n)$ in view of Remark \ref{rem:modifiedRVI} above. Compute iteratively
\begin{equation}
  x_i(n+1) = x_i(n) + b \left( y_{\zeta_i(n+1)}(n) - x_i(n) \right), \label{fast}
\end{equation}
where $\zeta_i(n+1)$ are chosen from the neighbors $\N(i)$ of $i$ in $\G$, with probability $r(i,j)$ independently of all else, and $b \gg a$. Then by the `two time scale' analysis of \cite[Sec.~9.3,~pp.~112-113]{BorkarBook}, $x(n) \approx \kappa^Ty(n)$ with high probability for large $n$, and we may use $x(n)$ as a surrogate for $y_{i_0}(n)$ in (\ref{RL}) with an asymptotically negligible error. We skip the details.

Our algorithm is scalable. To demonstrate scalability, a simulation was done on a randomly generated Erd\"{o}s-R\'{e}nyi graph of 100 nodes. The probability of an edge between a pair of nodes was 0.2. A symmetric matrix was then generated with each entry in the upper triangle having independent and uniform distribution over $[0,1]$ whenever a link exists between the corresponding nodes. Each row of this matrix was then normalized to sum to 1, and the resulting stochastic matrix was taken as $P$. The nodes were initialized with $x(0)$ generated with independent and uniform distribution over $[0,1]$. The plotted errors in Figure \ref{fig:ER-RVI-noise} are the supremum norms of errors $x(n) - (\eta^T x(0)) {\bf 1}$ and the subsequently time-averaged $z(n) - (\eta^T x(0)) {\bf 1}$. Data is noisy with AWGN variance 0.25. The updates were asynchronous, and the update rates differed across nodes. (Average inter-update time for node $j$ was $10+j$ time steps.) This difference was introduced in order to simulate the effect of activation set transmission constraints causing updates to occur at differing frequencies. This is a topic to which we now turn our attention.

\section{Gossip with CSMA}
\label{sec:gossip-csma}

Often in wireless systems not all links can be activated simultaneously. Asynchronous updates are therefore necessary. Let $\SA$ denote the collection of subsets of links $\E$ that can be {\em concurrently} active. The idea is to sample at each instant a subset $s$ from the permitted set $\SA$ according to some probability distribution $\varphi$ on $\SA$, and then activate all links in $s$. The probability distribution $\varphi$ should satisfy the constraints
\begin{eqnarray}
  \hspace*{-.1in} \lefteqn{ \sum_{s \in \SA}\varphi(s)I\{(i, j) \in s\} } \nonumber \\
  \hspace*{-.1in} &=& \hspace*{-.1in} \left( \sum_k \sum_{s \in \SA} \varphi(s) I\{(i,k) \in s\} \right) p(i, j) \ \forall \ (i, j) \in \E, \label{freq}
\end{eqnarray}
\begin{eqnarray}
  \hspace*{-.7in} \sum_{s \in \SA}\varphi(s) = 1, \mbox{ and } \varphi(s) \geq 0 \ \forall s \in \SA. \label{pi2}
\end{eqnarray}
The first constraint above ensures that link $(i,j)$ is activated with frequency $p(i, j)$. The next two constraints ensure that $\varphi$ is a probability distribution over $\SA$.

Any $\varphi$ that meets the constraints is a good candidate. But we shall choose the $\varphi$ that minimizes negative entropy
\begin{eqnarray}
  \mbox{Minimize} & & -H(\varphi) = \sum_{s \in \SA} \varphi(s) \ln \varphi(s) \label{eqn:Entropy-minimization}\\
  \mbox{subject to} & & (\ref{freq}) \mbox{ and } (\ref{pi2}). \nonumber
\end{eqnarray}
This choice of $\varphi$ leads to decentralization, yet respects the constraints imposed by the activation set $\mathcal{S}$, via use of a natural CSMA/CA strategy \cite{Jiang-Walrand}, \cite{Jiang}. We now describe the solution and its implementation.

The Lagrangian for problem (\ref{eqn:Entropy-minimization}), with Lagrange multipliers $\zeta = (\zeta_{i,j} \geq 0 , (i,j) \in \E)$ that relax (\ref{freq}), is
\begin{eqnarray*}
  \lefteqn{L(\varphi, \zeta) = - \ H(\varphi) } \\
   & & \quad \quad \quad  + ~ \sum_{i,j : i \neq j} \zeta_{i,j} \Big( p(i,j) \sum_k \sum_{s \in \mathcal{S}} \varphi(s) I \{ (i,k) \in s \} \\
   & & \quad \quad \quad \quad \quad \quad \quad \quad - ~ \sum_{s \in \mathcal{S}} \varphi(s) I \{ (i,j) \in s \} \Big).
\end{eqnarray*}
We now make the following remarks that naturally lead to the decentralized algorithm. Write $N_i(s) := \sum_{k \neq i} I \{ (i,k) \in s \}$.

The minimum of $L(\varphi, \zeta)$ over $\varphi$, subject to (\ref{pi2}), is easily seen to be
\begin{equation}
  \label{eqn:stationary-distribution}
  \varphi(s) = Z^{-1} \exp \left\{ \sum_{(i,j) \in s} \zeta_{i,j}  - \sum_i N_i(s) \sum_{k \neq i} p(i,k) \zeta_{i,k} \right\}
\end{equation}
where $Z$ is the normalization that makes $\varphi$ a probability distribution on $\mathcal{S}$. If $\zeta$ is constant, the following CSMA/CA strategy will realize (\ref{eqn:stationary-distribution}) as the stationary distribution of a continuous-time Markov chain. The CSMA/CA strategy (and the description of the Markov process) is the following:

\begin{itemize}
  \item[(a)] Let $s \in \mathcal{S}$ be the currently active set of links. For each inactive $(i,j) \in \mathcal{E}$, i.e., $(i,j) \in \mathcal{E} \setminus s$, if $s \cup \{ (i,j) \} \in \mathcal{S}$, then link $(i,j)$ changes state from inactive to active at rate $R_{i,j} = \exp\{\zeta_{i,j} - \sum_{k \neq i} p(i,k) \zeta_{i,k} \}$.
  \item[(b)] Each active link $v \in s$ changes from active to inactive at rate 1.
\end{itemize}

Note that these state changes can be done in a decentralized fashion. Let the inactive link be $u = (i,j)$. Node $i$ maintains Poisson clocks of rates $R_{i,j}$ for link $(i,j) \in \E$. When the earliest clock rings, say the clock for $(i,j)$ rings, node $i$ queries node $j$ if it can exchange packets, provided node $i$ sees no conflict. Node $j$ responds positively (acknowledgement or ACK) if there is no conflict, provides its information to node $i$, and node $i$ updates. If there is a conflict, node $i$ somehow gets to know this (e.g., no response from node $j$ or a negative acknowledgement (NACK)). If $(i,j)$ can become active, it does. Node $i$ then restarts all its clocks at appropriate rates $R_{i,k}$ (with $R_{i,j} = 1$ if $(i,j)$ is activated, and $R_{i,j} = \exp\{\zeta_{i,j} - \sum_{k \neq i} p(i,k) \zeta_{i,k} \}$ otherwise). We assume these exchanges are instantaneous. Carrier sensing is on all the time to check for conflicts.

The correct choice of Lagrange multipliers $\zeta$ that will solve the primal problem (\ref{eqn:Entropy-minimization}) is the maximizer of the Lagrangian with $\varphi$ taken to be (\ref{eqn:stationary-distribution}). To reach this maximizer, we shall locally update the Lagrange multipliers $\zeta(t)$ as a function of time $t$ at a slow time scale using stochastic gradient ascent on $L(\varphi, \zeta)$, where $\varphi$ is the current value. Observe that
\[
  \frac{\partial L}{\partial \zeta_{i,j}} = p(i,j) \sum_{s \in \mathcal{S}} \varphi(s) N_i(s)  - \sum_{s \in \mathcal{S}} \varphi(s) I \{ (i,j) \in s \}.
\]
This naturally suggests the following update equation for $\zeta(t)$. Let $T_k = k$ and let $\{\alpha(k)\}$ be decreasing positive weights that satisfy the conditions of Remark \ref{rem:StocApprox}. For $\sum_{k = 2}^{l} T_k = \theta_{l} \leq t < \theta_{l+1} = \sum_{k = 2}^{l+1} T_k$, set
\begin{eqnarray*}
  \lefteqn{ \zeta_{i,j}(t) = \zeta_{i,j}(\theta_{l-1}) } \\
  & & \hspace*{.4in} +~ \alpha(l) \left( p(i,j)(\theta_{l-1}) \#_i(l-1) - \#_{(i,j)}(l-1)   \right).
\end{eqnarray*}
Here, $\#_{(i,j)}(l-1)$ is the number of times link $(i,j)$ was active in the time period $[\theta_{l-1}, \theta_l)$ and $\#_i(l-1)$ is the number of times some link $(i, \cdot)$ was active in the same time period. We have replaced the gradient component
\[
  p(i,j) \sum_{s \in \mathcal{S}} \varphi(s) N_i(s) - \sum_{s \in \mathcal{S}} \varphi(s) I \{ (i,j) \in s \}
\]
with its stochastic variant
\[
  p(i,j)(\theta_{l-1}) \#_i(l-1) - \#_{(i,j)}(l-1),
\]
with averaging done over increasing durations. The quantity $\zeta$ is ascending along the direction of steepest ascent, and $\alpha(l)$ is a decreasing stepsize.

There are two technical issues vis-a-vis this iteration, which fortunately can be resolved in our favor. The first is that it is intended as a gradient ascent in the Lagrange multipliers to maximize the Lagrangian function. This it indeed is by the `envelope theorem' (see \cite[pp.~964-966]{Mas} or \cite[pp.~42-44]{Bardi} for a more general version). The other issue is that this is an asynchronous stochastic approximation. Nevertheless, the asymptotic behavior is still as desired as long as all components are updated `comparably often', a situation that holds true in our case, see, e.g., \cite[Ch.~7]{BorkarBook}.

\begin{figure}[tb]
\centering
\includegraphics[width=3.49in]{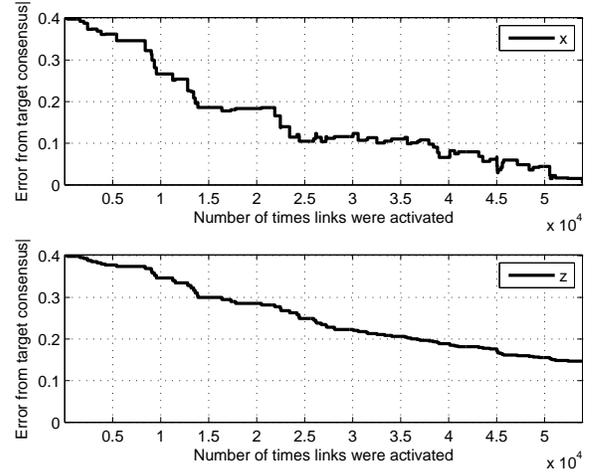}
\caption{Relative value iteration with CSMA/CA on an Erd\"os-R\'enyi graph with 100 nodes.}
\label{fig:ER-RVI-noise-CSMA}
\end{figure}

We illustrate the behavior of our relative value iteration scheme with CSMA/CA in Figure \ref{fig:ER-RVI-noise-CSMA} for a graph with 100 nodes. The set up is the same as in Figure \ref{fig:ER-RVI-noise}, except that the update rates are now governed by the CSMA/CA algorithm. The abscissa label is changed to `number of times links were activated', which is roughly two orders of magnitude smaller than `iteration number' for the parameters chosen. Naturally, comparing Figure \ref{fig:ER-RVI-noise-CSMA} with Figure \ref{fig:ER-RVI-noise}, we find that convergence is slow. This is the anticipated price paid for decentralization and learning of the Lagrange multipliers. An additional observation is that the maximum error can remain flat for long durations because of the slower time scale.

\section{Gossip miscellany}

\subsection{Better connectivity via multihop}
\label{subsec:multihop}

A further possibility to modulate the above scheme would be to use multihop. Consider, for example, the possibility of two hops. Let $j \in \V$ be a neighbor of $i \in \V$ and $k_1, k_2, \ldots, k_m \in \V$ be neighbors of $j$ that are not neighbors of $i$. Then each time $i$ polls $j$, $i$ may either pull the current value at $j$ or pull the value at some $k_{\ell}$ that has been already pulled and stored by $j$. Suppose the former is done with probability $p^0(i,j)$ and the latter with probability $p^{\ell}(i,j)$. This is tantamount to replacing the original $P$ by a modified $Q$ with additional edges from $i$ to the $k_{\ell}$'s with weights $p^{\ell}(i,j)$ resp., and replacing the weight $p(i,j)$ of edge $(i, j)$ by $p^0(i,j)$. There are, however, tradeoffs involved. For one, $i$ is actually sampling $j$'s value at a lower rate, thereby increasing the associated mean delay. Note also that the delay associated with $i$'s `virtual' sampling of $k_l$ will be a combination of delays due to its sampling of $j$ and $j$'s sampling of $k_l$. The benefit however is that the network is better connected.

We also need to choose the new sampling probabilities so as to retain the stationary distribution $\eta$, since our focus is on averaging with respect to $\eta$, not merely on obtaining a consensus. Some constraints suggest themselves, e.g., if $p(i,j,k)$ is the fraction of times $i$ polls $j$ in order to pull $j$'s stored value from $k$, then $\sum_{\ell'=0}^{m'} p^{\ell'}(j, k) \geq p(i, j, k)$, where $\ell'$ runs over the indices in $\{k_{\ell'}\}$ of neighbors of $k$, so as to avoid pulling the same value often. The tradeoffs and optimal choice of the parameters $p^{\ell}(i,j)$ are items for further study.

There is one simple, perhaps suboptimal, method that preserves the stationary distribution. Let $P$ be the given stochastic matrix. Node $i$ polls neighbor $j$ with probability $p(i,j)$. Once polled, node $i$ pulls node $j$'s value with probability $\alpha$, or pulls node $k$'s value from node $j$ with probability $(1-\alpha)p(j,k)$, where $k$ is a neighbor of $j$. The resulting stochastic matrix is $\alpha P + (1-\alpha) P^2$, instead of $P$, and its stationary distribution continues to be $\eta$.

\begin{figure}[tb]
\centering
\includegraphics[width=3.49in]{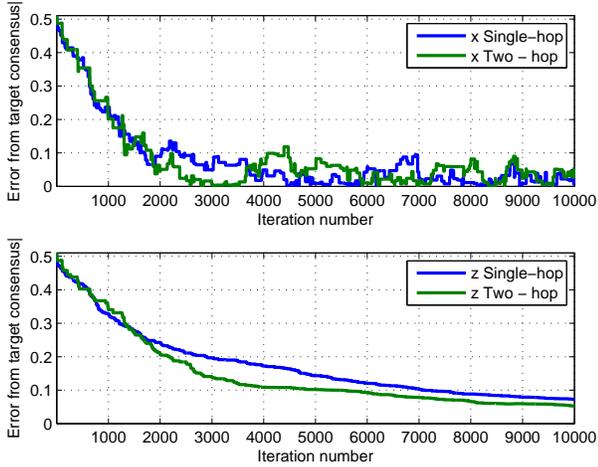}
\caption{Single-hop versus two-hop on an Erd\"{o}s-R\'{e}nyi graph of 100 nodes. In the top subplot, the curve that descends quickly and then fluctuates more is for two hops. In the bottom subplot, the bottom curve is for two hops.}
\label{fig:ER-single-hop-two-hop}
\end{figure}

See Figure \ref{fig:ER-single-hop-two-hop} for a comparison of single-hop with multihop. The setting is the same as in Figure \ref{fig:ER-RVI-noise}, but data is received without noise. The multihop curve used $\alpha = 0.8$. Note that in the simulation, stored data from a neighbor $k$ of $j$ who is also a neighbor of $i$ is occasionally pulled. With $\textsf{spec}(P)$ denoting the eigenvalues of $P$, the magnitude of the `second eigenvalue' of $\alpha P + (1 - \alpha) P^2$ satisfies
\[
  \max_{\lambda \in \textsf{spec}(P), \lambda \neq 1} |\alpha \lambda + (1 - \alpha) \lambda^2| < \max_{\lambda \in \textsf{spec}(P), \lambda \neq 1} |\lambda|,
\]
for $0 < \alpha < 1$. The improved spectral gap (from 1, 1 being the eigenvalue corresponding to the Perron-Frobenius eigenvector) implies faster convergence to correct consensus \cite[eqn.~(12.8)]{Levin-Peres-Wilmer}. The trade-off of delays notwithstanding, one does get faster convergence in Figure \ref{fig:ER-single-hop-two-hop}.

\subsection{Conditional importance sampling}
\label{subsec:importance-sampling}

\begin{figure}[tb]
\centering
\includegraphics[width=3.49in, height=3.49in]{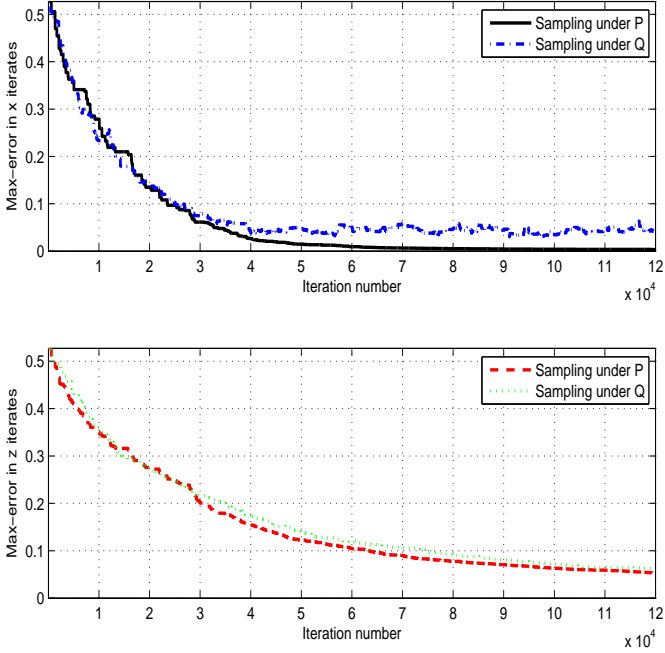}
\caption{Comparison of residual error under normal $P$ sampling and under importance sampling ($Q$) over time. Data is noiselessly received.}
\label{fig:nonoise-xzPQ}
\end{figure}

\begin{figure}[htb]
\centering
\includegraphics[width=3.49in, height=3.49in]{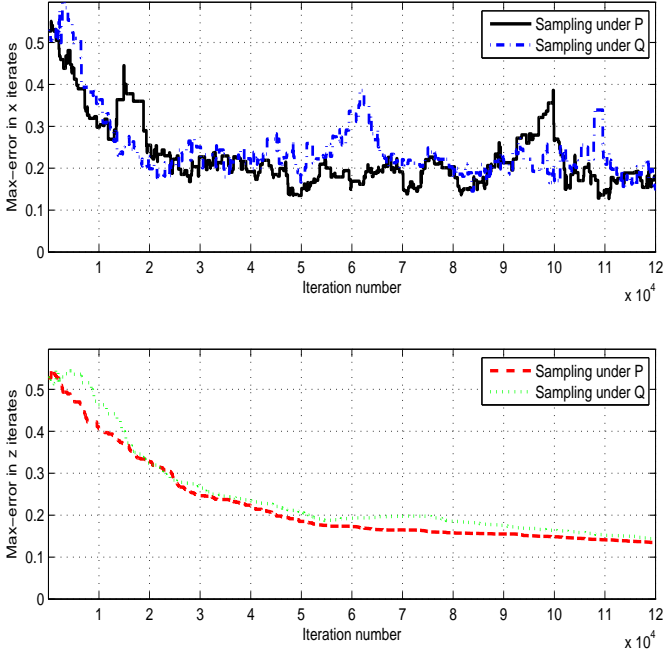}
\caption{Comparison of residual error under normal $P$ sampling and under importance sampling ($Q$) over time. Data is received with noise.}
\label{fig:noise-xzPQ}
\end{figure}

In the asynchronous case, there is an additional error term due to delays, over and above the errors due to discretization and noise. This can be shown to be bounded by a term proportional to $a$ and any bound on  the mean delays $\{E[\tau_{ji}(n)]\}$, where $\tau_{ji}(n)$ is the delay with which node $i$ received the value at node $j$ at time $n$. This is not surprising, because on the algorithm's time scale, in $\tau$ steps, each component would have changed by an amount that is $O(\tau)$. This suggests favoring low values of $E[\tau_{ji}(n)]$. Since this will be inversely proportional to the frequency with which $i$ polls $j$, one natural approach is to sample with a different polling matrix $Q := [[q(i, j)]]$ having
\begin{displaymath}
 q(i, j) > 0 \Longleftrightarrow p(i, j) > 0,
\end{displaymath}
and compensating for it by inserting the appropriate likelihood ratio correction as in \cite{Ahamed}. Thus we replace (\ref{sagossip}), (\ref{average}) by
\begin{eqnarray}
  x_i(n+1) \hspace*{-.05in} & = & \hspace*{-.05in} (1 - a(n+1,i))x_i(n) \nonumber \\
    & + &  a(n+1,i) \left( \ \frac{p(i, \xi_i(n+1))}{q(i, \xi_i(n+1))} \right) \cdot x_{\xi_i(n+1)}(n), \nonumber \\
    &   & \hspace*{0.75in} 1 \leq i \leq d, \ n \geq 0, \label{importance} \\
  z_i(n+1) \hspace*{-.05in} & = & \hspace*{-.05in} z_i(n) + \frac{1}{n+1}\left(x_i(n+1) - z_i(n)\right), \nonumber \\
    &   & \hspace*{0.75in} n \geq 0, \label{average-2}
\end{eqnarray}
where the stepsizes $a(n)$ are judiciously chosen so as to compensate for any differences in the relative rates of updates across nodes. (We omit the details of this compensation, but refer the reader to \cite[Sec.~9.3.(vi)]{BorkarBook}). In view of (\ref{importance}), the mean error due to delay in $i$ receiving $j$'s value is weighted by $q(i, j)\times\left(\frac{p(i, j)}{q(i, j)}\right) = p(i, j)$. The foregoing suggests choosing $Q$ to minimize
\begin{equation}
  \sum_i \sum_{j \in \N(i)} \frac{p(i, j)}{q(i, j)}, \label{cost}
\end{equation}
subject to the constraints $\sum_j q(i,j) = 1$ for every $i$ and $q(i,j) \geq 0$ for every $(i,j)$. It is easy to see that the optimal $Q$ is given by
\begin{displaymath}
  q(i,j) = \frac{\sqrt{p(i,j)}}{\sum_k \sqrt{p(i,k)}}.
\end{displaymath}
We tried this scheme as well, but the improvement for moderate sized problems was negligible, and the iterates are more noisy (as expected), suggesting that the $O(a)$ bound on delay errors is pessimistic. See Figures \ref{fig:nonoise-xzPQ} and \ref{fig:noise-xzPQ} for plots without and with noise, respectively, in the received data. Note that relative value iteration is not used. The simulation settings are the same as in Figures \ref{fig:ER-RVI-noise}, but with noiseless data in Figure \ref{fig:nonoise-xzPQ} and noisy data in Figure \ref{fig:noise-xzPQ} (AWGN noise variance 0.25). Despite the lack of improvement, we put this possibility on the table as it might prove useful for very large scale problems.

In the foregoing we used a constant stepsize $a > 0$. One could also use a slowly decreasing stepsize schedule $\{a(n)\}$ as in Remark \ref{rem:StocApprox}. The results will in fact be stronger: The conclusions of Theorem \ref{thm:RVI} can be strengthened to $y(n) \rightarrow V^*, y_{i_0}(n) \rightarrow \beta^*$ a.s. However, for the kind of applications that motivate this analysis, one is often in a nonstationary albeit slowly varying (compared to the algorithm's time scale) environment and the iterates are expected to track the slowly changing average. In such a scenario, the algorithm becomes too slow once $a(n)$ becomes very small, so it is preferred to use a small but constant stepsize $a > 0$. Again, there are trade-offs involved: small $a$ means lower fluctuations and a more graceful convergence, but slower speed, and large $a$ lead to faster convergence at the cost of higher variance. These are standard considerations for all stochastic approximation schemes and not specific to the ones under study here. For comparison, we present the analysis of the following sections under decreasing stepsizes.

\section{Estimating the Perron-Frobenius eigenvector}
\label{sec:pca}

In the previous problem, we sought a specific point on the ray defined by the Perron-Frobenius eigenvector of a stochastic matrix $P$, viz., $\bo{1}$. In many applications, see Section \ref{sec:applications} for some, one wants to identify the Perron-Frobenius eigenvector of a generic nonnegative matrix $Q$ modulo a scalar multiple. This is the problem we take up next, using gossip based computation.

Let $\G := (\V, \E)$ denote a connected graph with $\V, \E$ denoting respectively its node and edge sets. Nodes as before represent agents, $d := |\V|$, and we set $\V = \{1, \cdots, d\}$ without any loss of generality. Let $Q = [[q(i,j)]]_{i,j \in \V}$ denote an irreducible nonnegative matrix with Perron-Frobenius eigenvalue $\lambda > 0$ and the corresponding (Perron-Frobenius) eigenvector $q^* \in$ int$\left((\R^d)^+\right)$. Note that $q^*$ is specified as a unique vector in int$\left((\R^d)^+\right)$ only up to a positive scalar multiple. Consider the choice that satisfies
\[
  \alpha^T q^* = \sum_{i=1}^d\alpha_i q^*_i = \lambda
\]
for prescribed weights $\alpha = [\alpha_1, \ldots, \alpha_d]^T$, with all $\alpha_i > 0$, which renders $q^*$ unique. One scheme for computing $q^*$ is the iteration:
\begin{equation}
  \tilde{q}(n+1) = Q q(n), \ q(n + 1) = \frac{\tilde{q}(n+1)}{ \alpha^T \tilde{q}(n+1)}, \ n \geq 0. \label{power}
\end{equation}
This is immediately recognized as a variant of the `power method' \cite[Sec.~7.3]{Golub}. Alternatively, it is the special `linear' case of relative value iteration for risk-sensitive control of Markov decision processes \cite[Sec.~4]{BorkarMeyn}, except for a different normalization ($\alpha^T \tilde{q}(n+1)$ instead of $||\tilde{q}(n+1)||$).

The latter interpretation calls for a re-interpretation of $Q$ itself, which is the key to our `gossip' scheme. Let $\check{q}_i := \sum_{j=1}^d q(i,j), ~i \in \V$ and write $Q = DP$ where $P := [[p(i,j)]]_{i,j \in \V}$ is a stochastic matrix defined by
\[
  p(i,j) := \frac{q(i,j)}{\check{q}_i}, ~i \in \V,
\]
and $D :=$ diag$[\check{q}_1, \ldots, \check{q}_d]$ a diagonal matrix. Then $P$ is the transition matrix of an irreducible Markov chain $\{X(n)\}$ on $\V$ (irreducible because $Q$ is irreducible) and the $\check{q}_i$'s may be viewed as a per stage multiplicative cost. Then $\lambda$, thanks to  a standard multiplicative ergodic theorem \cite{Balaji}, equals the logarithm of the exponential growth rate of the multiplicative, or risk-sensitive, cost:
\begin{displaymath}
\lambda = \lim_{N\uparrow\infty}\frac{1}{N}\log E\left[\prod_{n=0}^{N-1}\check{q}_{X(n)}\right].
\end{displaymath}
Using this interpretation of $Q$, we consider the following gossip algorithm. Let $\{a(n)\}$ denote strictly positive stepsizes satisfying the conditions of Remark \ref{rem:StocApprox}. (We shall impose more conditions on $\{a(n)\}$ later as needed.)  At time $n + 1$, agent $i$ samples a `neighbor' $\xi_i(n+1)$ who is chosen according to the probability vector $p(i,\cdot)$ as before, independently of all else. Write $x(n) = [x_1(n), \ldots, x_d(n)]^T$. Agent $i$ then performs the computation:
\begin{eqnarray}
  \lefteqn {\nonumber x_i(n+1) = (1 - a(n))x_i(n) + a(n) \frac{\check{q}_i x_{\xi_i(n+1)}(n)}{\alpha^T x(n)} \label{firstcase}} \\
     &=& x_i(n) + a(n) \left[ \frac{\check{q}_i x_{\xi_i(n+1)}(n)}{\alpha^T x(n)} - x_i(n) \right] \nonumber \\
     &=& x_i(n) + a(n) \left[ \frac{\check{q}_i \sum_{j=1}^d p(i,j) x_j(n)}{\alpha^T x(n)} - x_i(n)\right] \nonumber \\
     &&  + \ a(n) M(n+1),
\end{eqnarray}
where $x_0 \in$ int$((\R^d)^+)$ and
\begin{displaymath}
  M(n) := [M_1(n), \cdots, M_d(n)]^T, n \geq 1,
\end{displaymath}
defined by
\begin{displaymath}
  M_i(n+1) := \frac{\check{q}_i \left( x_{\xi_i(n+1)} - \sum_{j=1}^d p(i,j) x_j(n) \right)}{\alpha^T x(n)}, n \geq 0,
\end{displaymath}
is a vector martingale difference sequence adapted to $\sigma$-fields $\F_n := \sigma(x(m), \xi_i(m), m \leq n, 1 \leq i \leq d), n \geq 0$. This leads to the vector iteration
\begin{equation}
   x(n+1) = x(n) + a(n) \left( \frac{D P x(n)}{\bar{x}(n)} - x(n) + M(n+1) \right), \label{iterate}
\end{equation}
where $\bar{x}(n) := \alpha^T x(n)$. We shall impose the convenient condition $\sum_i \alpha_i = 1$, though in what follows, this can be dropped with some extra work.

Equation (\ref{iterate}) is immediately recognized as another instance of a stochastic approximation algorithm, specifically a stochastic approximation counterpart of (\ref{power}). The term $\bar{x}(n)$ on the right hand side may seem to defeat the purpose of having a distributed scheme, but note that it can be separately computed by any distributed averaging scheme operating on a  faster time scale such as another `classical' gossip scheme as in (\ref{fast}).  This leads to the following result.

\begin{thm}
  \label{thm:convergence-power-iterations}
  The iterates $\{x(n)\}$ in (\ref{iterate}) remain almost surely bounded and converge almost surely to the eigenvector $q^*$ of $Q$ satisfying $\alpha^T q^* = \lambda$.
\end{thm}

\begin{IEEEproof} We provide only a sketch here because the proof follows the steps of \cite{200205MOR_Bor}. We associate with (\ref{iterate}) its `approximating o.d.e.'
\begin{equation}
  \dot{x}(t) = h(x(t)) := \frac{Q x(t)}{\bar{x}(t)} - x(t), \label{ode1}
\end{equation}
with $\bar{x}(t) := \alpha^T x(t)$ and $x(0) \in$ int$((\R^d)^+)$. We also consider another `limiting o.d.e.'
\begin{equation}
  \dot{\hat{x}}_t = h_{\infty}(\hat{x}_t), \label{ode2}
\end{equation}
where $h_{\infty}(x) := \lim_{a\uparrow\infty}\frac{h(ax)}{a} \ \forall x$. By Theorem \ref{appthm:bounded-sa} of Appendix,  $\{x(n)\}$ are a.s.\ bounded if (\ref{ode2}) has the origin as its unique asymptotically stable equilibrium. In our case, $h_{\infty}(x) = -x$, for which this is obvious. The first claim follows.

By arguments completely analogous to those in \cite{200205MOR_Bor} leading to \cite[Th.~4.1]{200205MOR_Bor}, we conclude that the positive orthant is an invariant set for (\ref{ode1}), and in this set, $q^*$ (with $\alpha^T q^* = \lambda$) is the unique asymptotically stable equilibrium for (\ref{ode1}). (We omit the details, which use the fact that the trajectories of (\ref{ode1}) are related to those of a secondary o.d.e.
\begin{equation}
  \dot{x}'(t) = \frac{Qx'(t)}{\lambda} - x'(t) \label{ode-sec}
\end{equation}
through a space-time scaling: $x(t) = \psi(t) x'(\tau(t))$ for suitably defined $\psi, \tau$ (see \cite[Lem.~4.1]{200205MOR_Bor})). It is also easy to see that the original iterates, if initiated in the positive orthant, remain in it, being successive convex combinations of elements of this convex set. In fact, one can argue as in \cite[Lem.~5.1]{200205MOR_Bor} to conclude that $\alpha^T x(n)$ remains uniformly bounded away from zero by a constant $\delta > 0$ that depends on the initial data. Therefore by confining to the appropriate subset of the positive orthant, we have Lipschitz property for the driving vector field of the o.d.e.\  (\ref{ode1}). Since  $q^*$ is the unique asymptotically stable equilibrium for (\ref{ode1}), the converse Liapunov theorem gives us a continuously differentiable Liapunov function for (\ref{ode1}). By Theorem \ref{appthm:convergence-sa} of Appendix, the second claim on convergence follows.
\end{IEEEproof}

We have used synchronous updates above, i.e., all components of the vector $x(n)$ are updated at the same time. If the scheme is totally asynchronous, i.e., each node autonomously polls neighbors at the instants of its choice, then one replaces the stepsize $a(n)$ above by $a(\nu(i,n))$ for the $i$th component, where `$n$' stands for some absolute global clock in the background and $\nu(i,n) :=$ the number of iterates executed by node $i$ till time $n$, a quantity known to $i$. Then under some additional conditions on the sequence $\{a(n)\}$ and $\{\nu(i,n)\}$ as in \cite[Sec.~7.4.(i)]{Borkar98}, it transpires that the limiting o.d.e.\ is simply a time-scaled version of (\ref{ode1}) and hence has the same asymptotic behavior. In particular, the above analysis applies. In fact, the aforementioned `global clock' may be a total artifice as long as causal relationships are respected.

\begin{thm}
 Assume that, for each $i$, the sequence $\nu(i,n)/n$ remains uniformly bounded away from 0. Then the conclusions of Theorem \ref{thm:convergence-power-iterations} continue to hold for the asynchronous iterates.
\end{thm}

\begin{IEEEproof}
The stability test used above for establishing a.s.\ boundedness of iterates has been extended to the asynchronous set-up in \cite{Bhatnagar}, which can be applied here to claim a.s.\ boundedness of iterates. Given that, we sketch here the convergence proof for the o.d.e.\ limits in the asynchronous case. Specifically, consider the pair of time-inhomogeneous o.d.e.s
\begin{eqnarray}
\dot{x}(t) &=& \Lambda(t)\left(\frac{Qx(t)}{\alpha^Tx(t)} - x(t)\right), \label{aode} \\
\dot{y}(t) &=& \Lambda(t)\left(\frac{Qy(t)}{\lambda} - y(t)\right). \label{bode}
\end{eqnarray}
Here $\Lambda(t)$ is a diagonal matrix with nonnegative entries on the diagonal. These o.d.e.s arise as counterparts of (\ref{ode1}), (\ref{ode-sec}) above in the analysis of the asynchronous scheme if one were to stick to the stepsize schedule $\{a(n)\}$ \cite{Borkar98}. Recall that $\nu(i,n)$ denotes the number of iterates of the $i$th component up to time $n$. Under the assumption that $\frac{\nu(i,n)}{n}$ remains uniformly bounded away from zero, the diagonal elements of $\Lambda(t)$ remain bounded from above and bounded away from zero from below \cite{Borkar98}. Intuitively, this means that all components are updated `comparably often', see \cite[Ch.~7]{BorkarBook}, for some sufficient conditions. Define
\begin{eqnarray*}
F(x) &:=& \frac{Qx}{\bar{x}}, \\
F_t(x) &:=& (I - \Lambda(t))x + \Lambda(t)F(x), \\
G(x) &:=& \frac{Qx}{\lambda}, \\
G_t(x) &:=& (I - \Lambda(t))x + \Lambda(t)G(x), \\
\|x\|^*  &:=& \max_i\frac{|x_i|}{w_i},
\end{eqnarray*}
where $w = [w_1, \ldots, w_d]^T$ is the right Perron-Frobenius eigenvector of $Q$ normalized to unit norm. Equations (\ref{aode}), (\ref{bode}) can be rewritten as
\begin{eqnarray}
\dot{x}(t) &=& F_t(x(t)) - x(t), \label{code} \\
\dot{y}(t) &=& G_t(y(t)) - y(t). \label{dode}
\end{eqnarray}
A direct verification shows that  $G(\cdot)$ is nonexpansive w.r.t.\ $\|\ \cdot \ \|^*$ and hence so is $G_t(\cdot)$. Furthermore, under our conditions on $\Lambda(\cdot)$, the set of fixed points for $G_t$ is exactly the same as that for $G$ for all $t \geq 0$. Hence the convergence of (\ref{bode}) to a positive scalar multiple of of the Perron-Frobenius eigenvector of $Q$  follows as in Theorem 4.2, p.\ 354, \cite{BoSo}. The rest of the argument closely mimics the proof of Lemmas 4.1 -- 4.2, pp.\ 301-303, of \cite{200205MOR_Bor}. The a.s.\ convergence of the algorithm then follows along the lines of Theorem~\ref{thm:convergence-power-iterations} above with minor modifications in the cited results of \cite[Ch.~7]{BorkarBook} to accommodate time-inhomogeneity.
\end{IEEEproof}

\begin{remark}
The modifications needed to the arguments of \cite{200205MOR_Bor} are quite routine. In fact, the equation in \cite{200205MOR_Bor} is much more general. There is a slight difference in our normalization by $\bar{x}(t)$ instead of by $x_{i_0}(t)$ for a fixed $i_0 \in \V$ as in \cite{200205MOR_Bor} which corresponds to the choice $\alpha_{i_0} = 1, \alpha_i = 0$ for $i \neq i_0$, but this calls for only a minor adaptation of the proof. Our choice is geared for the computability of the normalizing factor by another gossip algorithm as mentioned in the discussion following Remark \ref{rem:modifiedRVI}.
\end{remark}

We now present the finite time analysis of this algorithm, confining ourselves to the synchronous case of Theorem 3 for simplicity. This is much harder than in the previous case which involved linear iterates, because now we have a nonlinearity that is only locally Lipschitz. Just as in the convergence analysis above, our estimates here will depend on the initial data. To begin, we define as in \cite{200205MOR_Bor} the quantities
\begin{displaymath}
  \phi(n) := \max_i\left|\frac{x_i(n)}{q^*_i}\right|, \ \mu(n) := \min_i\left|\frac{x_i(n)}{q^*_i}\right|.
\end{displaymath}
Mimicking the arguments of Lemma 5.1 of \cite{200205MOR_Bor}, we obtain
\begin{displaymath}
  \phi(n+1) \leq (1 - a(n))\phi(n) + a(n) K,
\end{displaymath}
where $K := \frac{\phi(0)}{\mu(0)} \in (0, \infty)$. It follows that $\sup_n\phi(n) \leq K$ and thus, from the definition of $\phi(n)$ that
\begin{displaymath}
  \sup_n\|x(n)\| \leq \sqrt{d}K\|q^*\|_{\infty}.
\end{displaymath}
Note that this is a deterministic bound. Coupled with the already noted fact that $\alpha^T x(n) \geq \delta > 0 \ \forall n$, we have $\sup_n\|M(n)\| \leq C$ for a suitable \textit{deterministic} $C < \infty$. Thus Corollary 14, p.\ 43, of \cite{BorkarBook} can be applied to obtain an estimate of the type
\begin{displaymath}
  \Pr \Big(\|x(n) - q^*\| < \epsilon \ \forall \ n \geq n_0) \geq 1 - 2de^{-\frac{\hat{C} \epsilon^2}{\sum_{m \geq n_0}a(m)^2}}
\end{displaymath}
for $\epsilon > 0$, suitable $\hat{C} < \infty$, and $n_0$ exceeding a certain deterministic lower bound given by (4.1.4), p.\ 34, \cite{BorkarBook}. (The derivation of this estimate requires a continuously differentiable Liapunov function for (\ref{ode1}), which is guaranteed by the asymptotic stability of $q^*$ proved in Theorem 4.1 of \cite{200205MOR_Bor} and the converse Liapunov theorem.)

We now discuss some variants of the foregoing.

\subsubsection{A different normalization}
The following variant of (\ref{ode1}):
  \begin{equation}
    \dot{\tilde{x}}(t) =  Q \tilde{x}(t) - (\bar{\tilde{x}}(t)) \tilde{x}(t), \label{ode3}
  \end{equation}
and the associated iteration
  \begin{eqnarray*}
    \lefteqn{ x(n+1) = x(n) } \\
    & & + ~ a(n) \left( DPx(n) - (\bar{x}(n))x(n) + M'(n+1) \right)
  \end{eqnarray*}
for a suitably defined $\{M'(n)\}$, are also valid alternatives to (\ref{ode1}), (\ref{iterate}) resp. This is because (\ref{ode3}) is only a time-scaled version of (\ref{ode1}) and therefore has the same trajectories. Interestingly, (\ref{ode3}) also arises in a totally different context, viz., self-organization in complex systems \cite{Jain}.

\subsubsection{Better mixing}
Another variant in the spirit of the preceding remark is to use a transition matrix $\hat{P} = [[\hat{p}(i,j)]]_{i,j \in \V}$, with the property $p(i,j) > 0 \Longrightarrow \hat{p}(i,j) > 0$, in place of $P$ above. That is, sample $\xi_i(n+1)$ according to $\hat{p}(i,\cdot)$ instead of $p(i,\cdot)$ in the above. Further, replace (\ref{firstcase}) by
  \begin{eqnarray*}
    x_i(n+1) &=& (1 - a(n))x_i(n) + \\
      && a(n)\frac{\check{q}_i \ell(i,\xi_i(n+1)) x_{\xi_i(n+1)}(n)}{\bar{x}(n)},
  \end{eqnarray*}
where $\ell(i,j) := p(i,j) / \hat{p}(i,j)$ is the `one step likelihood function'. This scheme will have the same limiting o.d.e.\ as the original and hence the same asymptotic behavior. The matrix $\hat{P}$ can be chosen with an eye on the convergence rate, e.g., it can introduce transitions across quasi-invariant sets to reduce conductance and therefore facilitate mixing, even though such transitions have zero probability in the original transition matrix. Note that the one step likelihood for such transitions will be zero, thus `wasting' the learning step. Thus there is a trade-off involved.

\section{Applications}
\label{sec:applications}

As mentioned in the introduction, the primary application of spectral ranking  is in all kinds of evaluative exercises. In fact this field has a long history which, along with a fairly extensive list of applications, can be found in the excellent survey \cite{Vigna}. Here we touch upon a few instances of interest to us.

\subsection{Finding the argmin}

Let $\N(i)$ be a prescribed neighborhood of $i$ for each $i \in \V$ with $|\N(i)| \equiv N \ \forall i$ (for simplicity). Suppose we are given a function $\psi: \V \mapsto \R$ and want to find the point(s) at which it attains its minimum on $\V$, i.e., the $argmin$. Note that this is distinct from the problem of finding $\min \psi$. Consider the stochastic matrix $\check{P} = [[\check{p}(i,j)]]_{i,j\in \V}$ where for a prescribed temperature parameter $C > 0$,
\begin{eqnarray*}
    \check{p}(i,j) &=& 0, \  \notin \N(i), j \neq i\\
    &=& \frac{1}{N}e^{-\frac{(\psi(j) - \psi(i))^+}{C}},  \ j \in \N(i), \\
    &=& 1 - \sum_{k \neq i}\check{p}(i,k), \ j = i.
\end{eqnarray*}
Then its unique stationary distribution is the Gibbs distribution $\eta = [\eta_1, \cdots, \eta_d]^T$ for $\eta_i = Z^{-1}e^{-\frac{\psi(i)}{C}}, i \in \V$, where $Z$ is the normalizing factor. This concentrates on $argmin(\psi)$, peaking more and more as $C$ is lowered. In the above scheme, if we take $Q = \check{P}^T$, we get asymptotic convergence of the iterates in (\ref{iterate}) to $\eta$. One can in fact consider time-dependent $T(n)$ in place of $T$ that is decreased to zero at a suitable `cooling rate' as in simulated annealing \cite{Hajek}, so that $\pi(i)$ asymptotically concentrates (in probability) on  $argmin(\psi)$. See \cite{Anily} for more general variants of simulated annealing.

Observe that when $Q = \check{P}^T$ for a stochastic matrix $\check{P}$, $\bo{1}^TQ = \bo{1}^T$. Left-multiplying (\ref{ode1}) by $\bo{1}^T$, and taking $\alpha = \bo{1}$, we get
\begin{displaymath}
  \frac{d}{dt}(\bo{1}^T x(t)) = 1 - \bo{1}^T x(t),
\end{displaymath}
implying in particular that the simplex of probability vectors is invariant for this o.d.e.\ and attracts all trajectories starting away from it in the positive orthant. This raises the issue: why not drop the normalization by $\bar{x}(n)$ in the algorithm (\ref{iterate}), thus dropping the normalization by $\bar{x}(t)$ in (\ref{ode1}) and rendering it a linear system? More generally, why not use the prior knowledge of the Perron-Frobenius eigenvalue $\lambda$ when available, replacing $\alpha^T x(n)$ by $\lambda$ in (\ref{iterate})? The problem then is that the resultant linear o.d.e.\  converges to some point on the ray defined by $\{cq^*: c > 0\}$ which depends on where it starts, and the above stability test is inapplicable. Our earlier results on stochastic gossip indicate that convergence may still happen. Nevertheless, the above nonlinear scheme is superior because of its better stability properties: If $x$ is scaled by $a \gg 1$ , so are $\frac{Qx}{\lambda}$ and $-x$, but not $\frac{Qx}{\bar{x}}$, which remains unchanged. Thus as iterates blow up, the stabilizing negative drift is much more pronounced for the nonlinear scheme.

%Our simulations with a constant stepsize, not reported here, confirmed that this is indeed so. This problem, of course, will not arise in the corresponding deterministic iterates. However, some of the conventional fixes, such as projecting the iterates to a bounded set which has the driving vector field of the associated o.d.e.\ transversal to its boundary, will also help.

One can avoid taking the transpose by considering a `push' scenario instead of a `pull'. For simplicity, assume the fully asynchronous case when only one node polls her neighbor at a time. Suppose at time $n$, node $i$ polls node $j$ with probability $p(i,j)$ as before, but instead of `pulling' the value of $x_j(n)$ \textit{from} node $j$, it `pushes' the value of $x_i(n)$ \ita{to} node $j$, who then updates $x_j(n)$ according to
\begin{eqnarray*}
  \lefteqn{ x_j(n+1) =  x_j(n) + a(n) \ \times} \\
  &&  \left[ \frac{\sum_{i: j \in \N(i)}x_i(n) I\{\xi_i(n+1) = j\}}{\bar{x}(n)} - x_j(n) \right].
\end{eqnarray*}
If more than one node were to poll $j$ at the same time, then $j$ would have to do that many separate iterates at once.

Similar concerns arise for PageRank if one were to propose the above scheme for it, but the convergence issues there are much simpler (at least in theory). Evaluating PageRank \cite{Avra}, \cite{Berk}, \cite{Langville}, \cite{LangMe} in a distributed manner amounts to setting $Q =$ the transpose of the `Google matrix'. Note that this $Q$ is of the form $(1 - \epsilon)\tilde{Q} + \epsilon J$ where $\tilde{Q}$ is  transpose of a stochastic matrix and  $J := \frac{1}{d} {\bf 1} {\bf 1}^T$. Set $\alpha = {\bf 1}$. One can then write (\ref{iterate}) as
\begin{displaymath}
  x(n+1) = x(n) + a(n)\Big[(1 - \epsilon)\frac{\tilde{Q}x(n)}{\bar{x}(n)} + \frac{\epsilon}{d}\textbf{1} - x(n) + M(n+1) \Big].
\end{displaymath}
If one were to drop the $\bar{x}(n)$ in the denominator by invoking the prior knowledge of the Perron-Frobenius eigenvalue, then the right hand side is an affine function involving the matrix $(1 - \epsilon)\tilde{Q}$ which is a contraction with respect to the norm $\|x\|_1 := \sum_i|x_i|$. Thus convergence of the associated o.d.e.\ to its unique fixed point follows by \cite[Th.~4.2,~p.~354]{BoSo}.

\subsection{Other ranking problems}

A similar situation arises in the computation of eigenvector centrality, a centrality measure for social networks proposed by Bonacich (\cite{Bonacich}, see also \cite[p.~40]{Jackson}), for which $Q$ will be the column normalized adjacency matrix of the graph. Yet another instance is the `reputation networks' studied in \cite{Kamvar}, where the proposed `EigenTrust' algorithm is based on an analogous eigenvector computation.

Another ranking scheme is Kleinberg's `HITS' scheme which assigns two distinct rankings to nodes (the `\textit{hubs and authorities}' model), according to the Perron-Frobenius eigenvectors of $AA^T$ and $A^TA$ resp.,  $A$ being the adjacency matrix (\cite{Langville}, Chapter 11). The problem here is that both these matrices have possibly nonzero entries not only for neighbors, but also for neighbors' neighbors. But there is a simple fix for this. Instead of using a fixed matrix $Q$ as above, let $Q_1 = A$ and $Q_2 = A^T$ and alternate between the iterates:
\begin{eqnarray*}
  y_i(n) &=& \check{q}^1_i x_{\xi^1_i(n+1)}(n) \\
  x_i(n+1) &=&   x_i(n) + a(n) \times \\
    && \left[\frac{\check{q}^2_i y_{\xi^2_i(n+1)}(n)}{\bar{x}(n)} - x_i(n)\right],
\end{eqnarray*}
where $q^1_i, q^2_i$ are the row sums of the $i$th rows of $Q_1, Q_2$ resp., and the $\N(i)$-valued random variables $\xi^k_i(n+1), k = 1,2,$ are picked according to the probability distribution given by normalized $i$th row vectors of $Q_k, k = 1,2,$ resp. Then $x(n)$ converges to the Perron-Frobenius eigenvector of $A^TA$, almost surely. Interchanging the roles of $Q_1, Q_2$, would give the scheme for finding the Perron-Frobenius eigenvector of $AA^T$. The two are related through a simple linear transformation, so computing one of them suffices. (This is reminiscent of the `\ita{Randomized HITS}' scheme of \cite{Ng}.)

It may be noted that the Perron-Frobenius eigenvector of a positive definite matrix ($AA^T, A^TA$ in particular) could also be estimated by a pure Monte Carlo scheme based on the stationary distribution of a vertex (node) reinforced random walk \cite{Benaim}, \cite{Pemantle}, wherein a random walk revisits a state with a probability that is modified depending on the number of past visits. This, however, requires additional  book-keeping for each node, which may get cumbersome. For gradient--of--error based schemes for PageRank that use prior knowledge of the Perron-Frobenius eigenvalue, see \cite{Nazin}, \cite{NazPol}.

There are also other situations involving computation of the Perron-Frobenius eigenvector of a nonnegative matrix, e.g., in the regret based algorithms for learning in games (see Chapter 4 of \cite{Young}), where the above could be used as a `subroutine'. See \cite{Vigna} for even more examples.

\subsection{A reputation network}

Consider a social network of people with a common interest (books, movies, $\cdots$) who poll each other for their recommendations according to a stochastic matrix $P = [[p(i,j)]]$ and then rate the recommendation received (say, by $j$ from $i$ at time $n$) by a number $q(i,j)(n)$, e.g., on a scale $1-10$. The `reputation' of $i$ is carried by a number $x_i(n)$ which then is updated according to our algorithm given below in (\ref{eqn:reputation}). We have assumed here that only one person polls exactly one other person at any given time. More general scenarios can also be handled with some additional effort. The important things to note here are that: (i) this is a `push' model, and (ii) if $i$ samples $j$ with probability $p(i,j)$, then an `importance sampling' type adjustment is required to get the correct limiting behavior. The algorithm then is
\begin{eqnarray}
  \lefteqn {x_i(n + 1) = x_i(n) + a(n) \times} \nonumber \\
  \label{eqn:reputation}
  && \hspace*{.2in} \left[\frac{I\{\xi_i(n + 1) = j\} \ell(i,j)(n) x_j(n)}{\bar{x}(n)} - x_i(n)\right]
\end{eqnarray}
where $\ell(i,j)(n) := q(i,j)(n) / p(i,j)$.

The limiting o.d.e.\ is
\begin{displaymath}
  \dot{x}(t) = \frac{D Q x(t)}{\bar{x}(t)} - x(t),
\end{displaymath}
where $D := \mbox{diag}[\nu(1), \cdots, \nu(d)]$ with $\nu(i) :=$ the relative frequency with which node $i$ acts. This will then converge to the Perron-Frobenius eigenvector of $D Q$. If $D$ is a multiple of the identity matrix, this gives the Perron-Frobenius eigenvector of $Q$ which can be viewed as a relative reputation measure. More generally, the Perron-Frobenius eigenvector of $D Q$ implicitly incorporates a weight for the frequency of participation in the evaluation process by the node. This may be desirable, but if not, it can be avoided by explicitly compensating for the $\nu(i)$'s via a suitable choice of stepsizes as done in \cite[p.~87]{BorkarBook}.

\subsection{Identifying the principal eigenvector(s)}

We now consider an example where the aim is to find the principal eigenvector of a positive definite, though not necessarily nonnegative, matrix. This problem is inspired by \cite{Caramanis}. While it does not completely fit the above framework, similar techniques apply (not surprisingly, since power method does). Suppose we have $d$-dimensional random vectors $\{x(n), n \geq 1)$ that are i.i.d.\  $N(0, qq^T + \sigma^2I)$, where $\sigma > 0$ and $q \in \R^d$ is a fixed (but unknown) unit vector. On the generative side, this can be generated via $x(n) = q z(n) + w(n)$, where scalar random variables $\{z(n), n \geq 1 \}$) are i.i.d.\ $N(0, 1)$  and $\{w(n), n \geq 1\}$ are i.i.d.\ $N(0, \sigma^2I)$ vectors. The goal is to identify $q$ (the principal component). The power method (as per discussion with C.\ Caramanis; see simulations in \cite{Caramanis}) is as follows. Here $\{a(n)\}$ is a stepsize sequence satisfying the conditions of Remark \ref{rem:StocApprox}.

\begin{enumerate}
  \item Initialize $y(1) = x(1)$.
  \item For each $n \geq 1$ until a criterion to stop is reached (error less than a target or up to a maximum number of iterations), execute:
  \begin{eqnarray*}
    \lefteqn{ y(n+1) = (1 - a(n)) y(n) } \\
    & & \hspace*{.3in} +~ a(n) \Big\langle \frac{y(n)}{||y(n)||}, x(n+1) \Big\rangle x(n+1).
  \end{eqnarray*}
\end{enumerate}
Thus $\{y(n)\}$ will track the o.d.e.
\begin{eqnarray*}
  \dot{y}(t) &=& (qq^T + \sigma^2I)\frac{y(t)}{||y(t)||} - y(t) \\
  &=& A\left(\frac{y(t)}{||y(t)||}\right) - y(t),
\end{eqnarray*}
where $A = qq^T + \sigma^2I$. The solution $y(\cdot)$ converges to  the unique solution to $A\left(\frac{y}{||y||}\right) = y = ||y||\left(\frac{y}{||y||}\right)$, so that $||y||$ is the principal eigenvalue $\|q\|^2 + \sigma^2$ and $y$ is the principal eigenvector $(\|q\|^2 + \sigma^2)q$. We compare this method with the block-stochastic power method of \cite{Caramanis} which we describe next. Let $B \geq 1$ be the chosen block size. Let $T$ be the total number of samples. We take $T$ to be a multiple of $B$ for convenience. The block-stochastic power method is:
\begin{enumerate}
\item  Sample $Z \sim N(0, I)$, and initialize $z(0) = Z/||Z||$.

\item For each $\tau = 0, 1, . . . , \frac{T}{B} - 1,$ execute:
\begin{enumerate}
\item $\hat{z}(\tau + 1) = 0$.

\item For each $n = B\tau + 1, B\tau + 2, . . . , B\tau + B$, execute:
\begin{displaymath}
\hat{z}(\tau + 1)  = \hat{z}(\tau) + \frac{1}{B} \Big\langle z(\tau), x(n) \Big\rangle x(n).
\end{displaymath}

\item $z(\tau + 1) = \frac{\hat{z}(\tau + 1)}{||\hat{z}(\tau + 1)||}$.
\end{enumerate}
\end{enumerate}

\begin{figure}[tb]
\centering
\includegraphics[width=3.49in]{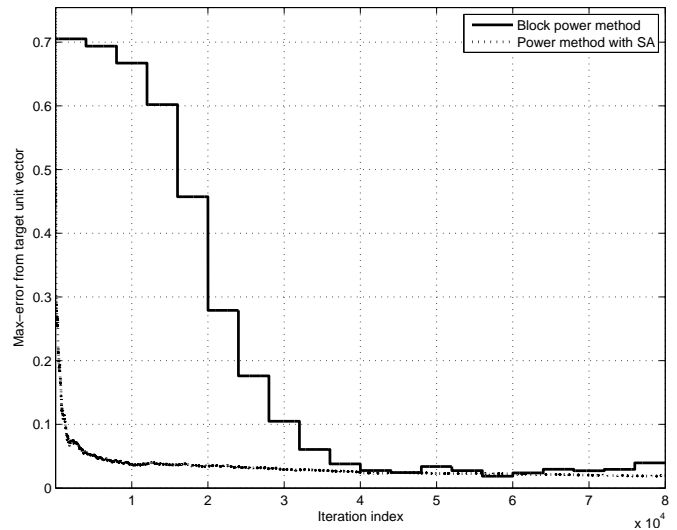}
\caption{Comparison of block stochastic with stochastic approximation.}
\label{fig:bssm}
\end{figure}

{\em Simulation results}: See Figure \ref{fig:bssm} for a comparison of block stochastic power method with stochastic approximation. The parameters for the simulation results are $d = 20, \sigma = 1, B = 4,000, T = 80,000$. This gives a total of 20 blocks for the block stochastic method. For the block stochastic method's error to be within 0.15 (with guarantees) \cite{Caramanis} suggests the conservative $T > 500,000$ with $B > 150,000$. Both methods have much faster convergence in practice. At the moment, we do not have better theoretical estimates on convergence rates or errors.

\appendices

\section{Stochastic approximation}
\label{appendix:StocApprox}

We recall relevant results from the theory of stochastic approximation. The archetypical stochastic approximation algorithm is the $d$-dimensional iteration
\begin{equation}\label{eq:BasicStocApproxAlgo}
  x(n + 1) = x(n) + a(n)[h(x(n), Y(n)) + M(n + 1)],
\end{equation}
with the following description.
\begin{itemize}
  \item $\{Y(n)\}$ is a `parametric Markov chain' on a finite state space $S$, i.e., there is a family of transition matrices $p_x = [[p_x(i, j)]]_{i,j \in S}, x \in \R^d,$ such that, for every $n \geq 0$,
  \begin{displaymath}
    \hspace*{-.05in} \Pr(Y(n+1) = j | Y(m), x(m), m \leq n) = p_{x(n)} (Y(n), j).
  \end{displaymath}
  For fixed $x$, $p_x$ is assumed to be irreducible with a (necessarily unique) stationary distribution $\nu_x$.
  \item $h: \R^d\times S \rightarrow \R^d$ is Lipschitz in its first argument.
  \item $\{a(n)\}_{n \geq 0}$ is a chosen positive stepsize sequence satisfying   $\sum_{n = 0}^{\infty} a(n) =\infty$ and  $\sum_{n =  0}^{\infty}(a(n))^{2}<\infty$.
  \item $\{M(n)\}$  is a  square-integrable martingale difference sequence w.r.t.\ the $\sigma-$fields $\{\mathcal{F}_n\},$  $$\mathcal{F}_n:=\sigma(x(0),M(i), Y(i), i \leq n),$$ satisfying $E[||M(n+1)||^{2}|\mathcal{F}_n]\leq L(1+||x(n)||^{2})$ a.s.\ for some $L >0.$
\end{itemize}
As $a(n) \rightarrow 0,$ (\ref{eq:BasicStocApproxAlgo}) can be viewed as a noisy discretization of the limiting ordinary differential equation (o.d.e.)
\begin{equation}
  \label{eq:limitingODE}
  \dot{x}(t) = \hat{h}(x(t)) := \sum_{i\in S}h(x(t), i)\nu_{x(t)}(i).
\end{equation}
This is the basis of the `o.d.e. approach' \cite{BorkarBook} for analyzing (\ref{eq:BasicStocApproxAlgo}) that we describe next. Assume:
\begin{description}
  \item[\bo{(A1)}] There is a continuously differentiable Liapunov function $V : \R^{n} \rightarrow [0,\infty)$ such that $\lim_{||x|| \rightarrow \infty} V(x) = \infty$, $\langle\nabla V(x), \hat{h}(x) \rangle < 0$ for $x \notin H:= \{x \in \mathbb{R}^n: \hat{h}(x) = 0\},$ which is assumed to be finite.
\end{description}

\begin{thm}
  \label{appthm:convergence-sa}
  Suppose $\sup_n \|x(n)\| < \infty$ almost surely. Then $\{x(n)\}$ converges almost surely to a possibly sample path dependent point in $H.$
\end{thm}

A test for almost sure boundedness is the following:
\begin{description}
  \item[\bo{(A2)}] For all $u$, $h_\infty(u) = \lim_{c \uparrow \infty}(\sum_ih(cu, i)\nu_{cu}(i))/c$ exists ($h_{\infty}$ will be necessarily Lipschitz) and the o.d.e. $\dot{x}(t)=h_{\infty}(x(t))$ has origin as its globally asymptotically stable equilibrium.
\end{description}

\begin{thm}
\label{appthm:bounded-sa}
Under $\bo{(A2)}$, $\sup_n\|x(n)\| < \infty$ almost surely.
\end{thm}

Theorems \ref{appthm:convergence-sa} and \ref{appthm:bounded-sa} follow from Theorem \cite[Th.~7,\ p.\ 74]{BorkarBook}, and Theorem \cite[Th.~9,\ p.\ 75]{BorkarBook}.

The `constant stepsize' version of the iteration uses $a(n) \equiv a > 0$. Here the above statements have weaker counterparts as follows:

\begin{thm}
\label{appthm:convergence-constant-a}
Suppose that $\sup_n E[\|x(n)\|^2] < \infty$.  Then $\limsup_{n \uparrow \infty} E[\inf_{y \in H}\|x(n) - y\|^2] = O(a)$.
\end{thm}

\begin{thm}
\label{appthm:bounded-constant-a}
Under $\bo{(A2)}$, $\sup_nE[\|x(n)\|^2] < \infty$.
\end{thm}

For details, see \cite[Ch.~9]{BorkarBook}.

%\section*{Acknowledgment}
%
%Portions of this work were presented at the 50th Allerton Conf.\ on Communication, Control, and Computing, Oct.\ 1-5, 2012, Monticello, IL, \cite{bormak} and Workshop on Info.\ Theory and Appl., San Diego, Feb.\ 10-15, 2013 \cite{bormaksun}. Research of V.\ S.\ Borkar was supported in part by a J.\ C.\ Bose Fellowship and a grant ``Distributed Computation for Optimization over Large Networks and High Dimensional Data Analysis'' from Department of Science and Technology, Government of India. R.\ Sundaresan was supported by the Indo-US Science and Technology Forum Fellowship and by the US National Science Foundation under grant CCF-1017430. This author thanks the Coordinated Sciences Laboratory, University of Illinois at Urbana-Champaign, for its hospitality during the course of this work.

\end{document}